\documentclass[12pt, draftclsnofoot, onecolumn]{IEEEtran}
\usepackage{graphicx} 
\usepackage{xcolor}
\usepackage{booktabs}
\usepackage{amsmath}
\usepackage{cite}

\title{Modeling and Analysis of Crypto-Backed Over-Collateralized Stable Derivatives in DeFi}

\author{\IEEEauthorblockN{Zhenbang Feng, Hardhik Mohanty, and Bhaskar Krishnamachari}\\
\IEEEauthorblockA{\textit{Department of Electrical and Computer Engineering} \\
\textit{Viterbi School of Engineering} \\
\textit{University of Southern California}\\ Los Angeles, CA, USA \\
\{jasonfen, hmohanty, bkrishna\}@usc.edu}}

\date{February 2024}

\begin{document} 

\maketitle

\begin{abstract}
    In decentralized finance (DeFi), stablecoins like DAI are designed to offer a stable value amidst the fluctuating nature of cryptocurrencies. 
    We examine the class of crypto-backed stable derivatives, with a focus on mechanisms for price stabilization, which is exemplified by the well-known stablecoin DAI from MakerDAO. For simplicity, we focus on a single-collateral setting. We introduce a belief parameter to the simulation model of DAI in a previous work (DAISIM), reflecting market sentiments about the value and stability of DAI, and show that it better matches the expected behavior when this parameter is set to a sufficiently high value. We also propose a simple mathematical model of DAI price to explain its stability and dependency on ETH price. Finally, we analyze possible risk factors associated with these stable derivatives to provide valuable insights for stakeholders in the DeFi ecosystem.

\end{abstract}

\section{Introduction}
In the modern decentralized finance (DeFi) ecosystem, stablecoins have emerged to counter the volatility inherent to cryptocurrencies like Bitcoin or Ethereum \cite{kolodziejczyk2020stablecoin}. Pegged to stable reserves such as fiat currencies, precious metals, or a diversified portfolio of assets, stablecoins try to maintain a consistent value. These stable derivatives are mechanisms for hedging risks, speculating on future price movements, and improving capital efficiency \cite{ante2021influence}. The stability of these tokens is maintained using a range of strategies, such as backing by fiat currencies, crypto-assets, or the use of algorithmic techniques that drop the need for conventional asset collateral. Such innovations in stabilization techniques highlight the versatility and adaptability of stablecoins in the digital age, reflecting a significant step forward in the quest for stability in the dynamic landscape of DeFi.

Among the plethora of stablecoins, DAI from MakerDAO stands out as a flagship stable derivative designed to maintain a steady value relative to \$1 (USD). Initially released as a single-collateral derivative known as SAI, it has evolved into a more sophisticated multi-collateral format to enhance stability and resilience. Expanding upon this secure foundation, the transition from SAI to DAI brought about the capacity to accept multiple types of cryptocurrency as collateral, not just Ethereum (ETH). This diversification of backing assets significantly enhances the stability of DAI. The MakerDAO protocol allows users to mint SAI by creating collateralized debt positions (CDPs), which are essential components of the system \cite{team2020maker}. In these smart contract-driven constructs, users must deposit ETH as collateral, over-collateralizing to accommodate for volatility in order to produce SAI and maintain its peg to \$1.

Once minted, DAI serves as a versatile tool, functioning as a medium of exchange, a store of value, and a unit of account. The protocol also provides users an opportunity to earn interest on their DAI holdings through the DAI Savings Rate. DAI's transparent and decentralized governance exercised by MKR token holders ensures that it remains stable. MKR token holders cast votes on vital protocol decisions, such as which assets qualify as collateral and the risk parameters for these assets. Furthermore, the transparent recording of every transaction, voting on the blockchain, and DAI's adoption in over 400 different apps and services spanning wallets, DeFi platforms, and games reinforce its position as a reliable and integral component of the modern cryptocurrency landscape.

Research on stablecoins is still emerging, with a few key studies paving the way for our investigation. Lyons \textit{et al.} \cite{lyons2023keeps} provided a foundational understanding of the functional efficiency of stablecoins by scrutinizing the mechanisms through which they maintain their peg in the digital economy. Building on this, Mita \textit{et al.} \cite{mita2019stablecoin} categorized the stablecoins based on the nature of their collateral and analyzed them through the lens of traditional economic models such as Hayek money and Tobin tax. Furthermore, Mundt \textit{et al.} \cite{klages2020stablecoins} delved into noncustodial stablecoins, revealing how liquidation protocols and investor behavior shape their stability, particularly in tumultuous market conditions.
Complementing these theoretical perspectives, Gudgeon \textit{et al.} \cite{gudgeon2020decentralized} critically assessed the security of governance in platforms like MakerDAO, a key aspect of stablecoin infrastructure. On a more practical note, Kothari \textit{et al.} \cite{kothari2018simulating} utilized simulations to offer insight into the demand dynamics for asset-backed stablecoins amidst external price shocks, a crucial factor for understanding market resilience. Finally, Clark \textit{et al.} \cite{clark2020demystifying} provided a broad survey that contextualizes the stablecoin landscape, setting the context for more detailed analysis. Drawing from these insights, our work aims to add to the understanding of DAI within DeFi, especially how it maintains its value in various economic conditions.

Synthetix and Mirror Protocol represent pivotal advancements in blockchain-based synthetic derivatives, both drawing on mechanisms similar to the DAI stablecoin from MakerDAO for maintaining price stability. 
Synthetix \cite{synthetix} allows users to create synthetic assets known as Synths, which track the value of real-world assets like fiat currencies, cryptocurrencies, and cryptocurrency indexes, and commodities such as gold and silver. The primary mechanism for minting Synths in Synthetix involves over-collateralization with the native cryptocurrency of the platform, SNX. The minted Synths are backed by a 600\% collateralization ratio determined by community governance. Synthetix uses a combination of staking rewards and exchange fees to incentivize users to maintain the peg of Synths. The platform relies on decentralized oracles to provide accurate and real-time price feeds of the underlying assets.
Mirror Protocol \cite{mirror} operates on the Terra blockchain and enables the creation of synthetic assets called Mirrored Assets (mAssets). These crypto tokens are utilized to track the price of real-world assets such as stocks. Mirror Protocol primarily uses Terra stablecoins like UST to mint mAssests with 150\% over-collateralization. The value of mAssets is maintained through a combination of minting liquidations, arbitrage, and governance. Mirror also relies on decentralized price oracles to provide real-time data on the prices of the underlying real-world assets.
Both platforms exemplify the synergy between cryptocurrency innovation and traditional financial stability, mirroring DAI’s approach of using over-collateralization and real-time data to maintain a stable asset value.

This paper sets out to enhance the simulation models for DAI stablecoin, particularly focusing on the associated risk factors and the efficacy of its price stabilization mechanisms. Utilizing empirical data and employing a variety of modeling methods, we plan to delve into the complex factors that influence the value of the DAI peg. Building on previous work, DAISIM \cite{bhat2021daisim}, we consider the impact of beliefs and narratives, which appear to play a significant role in DAI's valuation alongside the more technical aspects of its protocol. Our analysis aims to shed light on the nuanced interaction between market behavior and the design of the protocol that together maintains DAI's stability. In particular, our contributions can be listed as follows:
\begin{itemize}
    \item We introduce a belief parameter that captures market sentiment about DAI's valuation and stability in the modeling of the DAI stablecoin used in DAISIM. We show that setting this parameter to a sufficiently high value results in the DAI converging to a value close to \$1 (USD). 
    \item We present a mathematical model that captures the stability mechanism of DAI in relation to ETH price fluctuations. This model offers a quantitative understanding of the influence of ETH price on the behavior of DAI stablecoin under different market conditions.
    \item We provide a risk analysis for stable derivatives. This includes examining the impacts of various factors, such as oracle reliability, debt ceiling, and smart contract vulnerabilities, on the stability and operation of these stablecoins.
\end{itemize}

The remainder of this paper is organized as follows: Section \ref{sec2} offers a detailed analysis of the DAI mechanism, its structure, user interaction, and the role of price oracles. Section \ref{sec3} discusses the mechanisms for price stabilization of DAI. Section \ref{sec4} introduces DAISIM and our revised objective function. Section \ref{sec5} presents the simplified theoretical foundation of this model. Section \ref{sec6} explores the relationship between collateral price and derivative price using empirical data and simulations. Section \ref{sec7} examines various risk factors in stable derivatives. The paper concludes with Section \ref{sec8} summarizing our findings and insights.

\section{The Mechanism} \label{sec2}
Our particular interest lies in the mechanisms that enable these synthetic assets to maintain track of their underlying assets under market pressures and volatility. MakerDAO's DAI stands out as an important example of such a synthetic derivative, a stablecoin designed to peg US Dollar. We find it instructive to explore the inner workings of DAI's stability protocol, which relies mostly on decentralized mechanisms without centralized authority. While DAI has transitioned to a multi-collateral DAI starting in November 2019, our analysis below will delve into the single collateral form, known as SAI, which operated from December 18, 2017, until the protocol's transition to the multi-collateral system. SAI reflects the core mechanism of DAI and provides ease of exposition and analysis.
\subsection{Basic Structure and Functionality of DAI}
DAI functions by allowing users to lock up collateral (e.g., Ethereum) in a smart contract. The action of doing this is called opening a Collateralized Debt Position (CDP).  The process of collateral lock-up and DAI generation was designed to ensure over-collateralization, typically requiring a collateral value exceeding the value of DAI by at least 150\%. This requirement ensures the creation of a buffer to absorb market volatility. The specific over-collateralization ratios are subject to change based on governance decisions and market dynamics throughout SAI's tenure.

\subsection{User Interaction with DAI}
\begin{figure}[h!]
    \centering
    \includegraphics[width=0.8\linewidth]{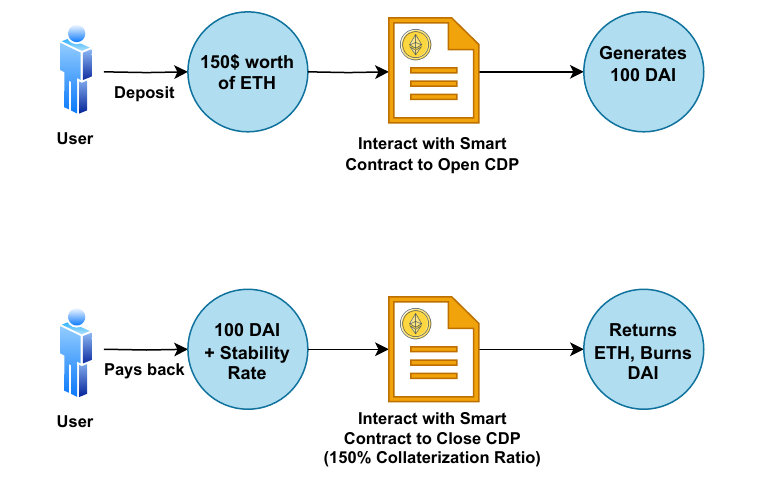}
    \caption{Illustrations of opening and closing CDP}
    \label{fig:dai_user}
\end{figure}

Users typically interact with DAI in the following ways. As illustrated in Figure-\ref{fig:dai_user}, users generate a CDP and obtain DAI by locking up their collateral assets. This DAI can then be used to transact, invest, or leverage in other decentralized financial applications. When users wish to retrieve their collateral, they repay the DAI they generated plus a stability fee to unlock their original assets. In addition to direct interaction with CDPs, users can buy or sell DAI on various exchange platforms just like any other cryptocurrency.

\subsection{Role of Price Oracles}
Oracles serve as bridges that bring real-world information, in this case market data, into the blockchain environment. In the context of the DAI stablecoin mechanism, oracles play an indispensable role, because DAI's stability and over-collateralization rely on the accurate and timely valuation of collateral assets. Oracles provide this service by feeding external data, in this case, the current market prices of collateral assets, into the blockchain. In DAI, this data is used to verify whether the value of the collateral in a CDP falls below a certain threshold, and if so, triggers a liquidation event to ensure DAI remains adequately backed. 


\subsection{The Liquidation Mechanism}
While the emphasis often rests on creation and utilization of DAI, an equally important mechanism is how the system handles potential under-collateralization. If the value of the collateral in a CDP drops below a predefined threshold, the system automatically liquidates a portion of the collateral to ensure DAI remains over-collateralized. This liquidation process involves auctioning off the collateral for DAI. Keepers, independent actors in the ecosystem, play a significant role in these auctions, bidding on and buying the liquidated collateral. This automatic and decentralized liquidation process is essential to maintain trust in DAI's stability, even during significant market downturns.

\section{Price-Stabilizing Mechanisms} \label{sec3}
Synthetic derivatives like Single-Collateral DAI (SAI) aim to mirror a particular asset's value. In SAI's case, it seeks to peg its value to the US Dollar. The question then arises: How does it ensure that its value remains stable at 1 USD? In reality, several mechanisms work together to ensure this stability.

\subsection{Emergency Shutdown}
Emergency shutdown is a mechanism designed for extreme circumstances. This process can be triggered during events like security breaches or huge market volatility. It allows DAI holders to directly redeem its value for \$1 of collateral at the point in time when the Emergency Shutdown is initiated, essentially enforcing the 1:1 USD soft peg.


\subsection{DAI Savings Rate or Stability Fees} 
Another significant mechanism to ensure price stability is the DAI Savings Rate. When market dynamics push the DAI price away from the target, this savings rate is adjusted. If DAI's market price goes above 1 USD, the savings rate decreases, reducing demand and pushing the price towards 1 USD. Conversely, if DAI's price falls below 1 USD, the savings rate rises, increasing demand and nudging the price upwards. In the Single-Collateral DAI, there isn't a DAI Savings Rate; instead, stability Fees on CDPs are adjusted directly to balance the demand and supply.

\subsection{Keepers}
Keepers play an invaluable role in maintaining DAI's stability. These are typically automated agents incentivized by profit. They actively participate in Debt and Collateral Auctions during CDP liquidations. More crucially, they trade DAI around its target price, selling when the price is high and buying when it's low, banking on the long-term convergence towards the target price, thereby helping in maintaining the price stability. Individuals or entities are often motivated to operate keepers due to the potential for profit. This profit expectation relies on their belief that the price of DAI will remain stable around 1 USD.

\subsection{Narrative or Belief}
An often underestimated element in the mechanism of stability is the narrative or the shared belief of its participants. While tangible mechanisms like target prices, keepers, and oracles play vital roles, the intangible narrative equally underpins the stability of such systems.

For example, the shared belief in DAI(SAI) is that it will stay aligned with the US dollar. This belief is propagated and reinforced through consistent communication, historical performance, community trust, and governance transparency. When participants uniformly believe in the 1:1 pegging, they act in ways that naturally enforce this peg further. For instance, if the price of SAI drifts slightly above 1 USD, the narrative would drive participants who are believing to sell, thereby pushing the price back to the peg. Conversely, if SAI dips below 1 USD, the narrative would deem it undervalued, triggering buys and bringing the price back to one dollar.

This shared narrative isn't solely built on blind faith. When the system consistently demonstrates its ability to maintain the peg through various market conditions, it reinforces the narrative. In the following sections, we will delve deeper into the role of this narrative by simulating the effects of shared belief.

\subsubsection{Role of Belief on DAI price dynamics}
Let's consider the DAI to be solely influenced by ``belief" (when $b$ is high). It becomes apparent that a shared consensus on price among the majority of investors creates a situation where an individual investor with a different view is likely to experience a negative expected return. Thus, an investor without extra information would tend to follow the majority, which again reinforces the consensus. Let's consider two illustrative scenarios:
\begin{itemize}
    \item First, suppose the current price of DAI is 1.2 USD. If investor Bob believes that the price will rise to 1.5 USD, he would purchase or take a long position in DAI, while the majority, anticipating a price decline, would sell or short DAI. 
 \item Second, consider a situation where the price of DAI is 0.8 USD. If Bob expects the price to drop further to 0.5 USD, he would choose to sell or short DAI. However, if the majority believes the price will increase, they would be looking to buy or go long on DAI.
\end{itemize}

In essence, any given price could theoretically serve as a Nash Equilibrium. However, the mechanism of over-collateralization, Emergency Shutdown, and narrative set by MakerDAO suggest that the Schelling point is anchored at 1 USD.

\section{Simulating DAI} \label{sec4}

\subsection{DAISIM}


DAISIM \cite{bhat2021daisim} presents a detailed computational simulation of the single-collateral DAI stablecoin, which was initiated by the MakerDAO project in 2017. This study places a significant emphasis on modeling the behavior of cryptocurrency investors, who are characterized by a range of risk tolerances and are considered rational in their portfolio optimization strategies. A key feature of this simulator is its incorporation of automated mechanisms for both order matching and updating prices, which are critical for determining the market price of DAI. The primary aim of this research is to investigate how the price of single-collateral DAI and the allocation of investment portfolios vary among a group of investors. This variation is analyzed in the context of exogenous parameters such as the price of ETH, as well as several internal system parameters, including the stability rate and transaction fees.

In terms of the simulation's structure, the system parameters play a pivotal role in shaping its dynamics. These include the Stability Rate, represented as $r_s$, and the Transaction Fees, denoted as $\beta$. The simulation also considers the size of the market, denoted as $n$, which represents the number of investors involved. Each investor within this simulation is depicted with a distinct risk profile, defined by a weight parameter $\lambda_i$ for the $i^{th}$ investor. Additional parameters are related to the price update algorithm used in the simulation.
In the simulated market environment created by DAISIM, investors have the opportunity to engage in a variety of transactions. They can buy ETH at the current market price, represented by $P_{ETH}$, and have the ability to open and close Collateralized Debt Positions (CDPs) in line with the current stability rate. Additionally, they can engage in the buying and selling of DAI within the simulated marketplace. Importantly, all these transactions incur a constant transaction fee, as specified by $\beta$. A key function of the simulator is to match buy and sell orders for DAI and to determine the market-clearing or settling price for DAI, denoted by $P_{DAI}$.


\subsection{Updated objective function}
The initial objective function proposed in DAISIM model centered around the asset optimization mechanism. It used a vector \( x = [x_{i,1}, x_{i,2}, x_{i,3}, x_{i, 4}] \) to denote holdings in USD, ETH, DAI, and cETH. Considering the single collateral DAI, the objective function to be maximized was originally formulated in DAISIM as follows:
\begin{equation}
x^T \mu - \xi x^T \Sigma x - \frac{x_{i;4}}{\rho} r_s - \tau
\end{equation}
However, the simulated price of the DAI driven by this objective function displayed significant deviation from the desired 1 USD peg. Such consistent deviations suggest that external factors, potentially rooted in market beliefs about DAI's stability were absent from the model's considerations. Therefore, we introduce ``belief factor'' ($\beta$) to better encapsulate market sentiment regarding DAI's valuation. 
Incorporating this belief factor accounts for the dual nature of market reactions to DAI's price fluctuations. When the price of DAI falls below \$1, the positive $\beta$ terms stimulate an increase in demand and a decrease in supply, aiming to correct the price towards its intended \$1 peg. Conversely, when DAI's price exceeds \$1, these terms adjust to increase supply and reduce demand, facilitating a return to the pegged price. When integrated into the model dynamics, these adjustment mechanisms highlight the investor's expectation that DAI should maintain a value of \$1.   
Hence, the refined objective function to be maximized can be written in the following manner:
\begin{multline}
x^T \mu - \xi x^T \Sigma x - \frac{x_{i;4}}{\rho} r_s - \tau + \beta \left( \frac{x_{i;3}}{P_{DAI}} (P_{DAI} - 1) \right) - \beta \left( \frac{x_{i;4}}{P_{DAI}} (P_{DAI} - 1) \right)
\end{multline}
The term \( x_{i;3} \) serves as a representation of the market's buying and selling activities concerning DAI. When demand for DAI escalates, there's a natural inclination for its price to surge. Conversely, excessive selling or an oversupply can pressure its price downward. The term's design, especially when modulated by the ``belief factor'' ensures that the objective function is either rewarded or penalized based on DAI's price alignment with its 1 USD peg. On the other hand, collateralized ETH (\( x_{i;4} \)) relates to the dynamics of DAI minting. It signifies the amount of DAI minted by locking up collateral, typically ETH, in the context of single collateral DAI. The act of minting directly influences DAI's market supply. An unchecked increase in supply, if not met with corresponding demand, can tilt the scales, causing DAI's price to drop. Conversely, reducing DAI's supply by burning or destroying it can have the opposite effect. This term's inclusion in the objective function ensures that such supply adjustments are swiftly addressed, especially when they deviate DAI's price from its intended peg. The modified code is available on the DAISIM 2.0 github repository \cite{DAISIM2}.

\section{Simplified mathematical model for DAI} \label{sec5}

Here we present a simple theoretical model of DAI price. This model is presented primarily as an explanatory tool - we do not claim that this is the ``correct" model of reality. However, this model allows us to explain two phenomena: 1) how belief influences the price being close to 1, reflected in a high demand for DAI at price below that value and a low demand beyond that price, and 2) how dependence of supply and demand for DAI on the price of ETH might impact the relationship between ETH price and the valuation of DAI. This model could potentially serve as a baseline for further investigation and development by the research community.

First, we model the supply of DAI as linearly increasing in its price  $P_{DAI}$ as well as in the price of ETH $P_{ETH}$. And we assume that the stability rate $\gamma$ discounts the price of the locked ETH as $\frac{P_{ETH}}{1+\gamma}$. This captures the intuition that when the stability rate is higher, a user is dissuaded from opening a CDP and creating more supply of DAI. This gives us that supply $S_{DAI}$ is given by: 

\begin{equation}
    S_{DAI} = k \cdot \frac{P_{ETH}}{(1+\gamma)} P_{DAI}
\end{equation}

where $k$ is just a constant of proportionality. 

Likewise, we can model the demand $D_{DAI}$ as a linearly decreasing function of $P_{DAI}$. If we ignore other terms for now, $m$ and $c$ represent the slope and intercept of the line. But we also introduce a variable $b$ to capture the belief in the price of DAI being equal to 1 USD. When $b$ gets larger, the demand decreases sharply from a high value to a low value around the price point $P_{DAI}$ = 1 USD. Finally, the term $\alpha P_{ETH}$ creates a proportional relationship between the demand for DAI and the price of ETH, reflecting, for example, that there may be more demand in using DAI during a bull market run. 

\begin{equation}
    D_{DAI} = - (m + b - \alpha P_{ETH}) P_{DAI} +  (b + c)
\end{equation}

To determine the price, we equate these two as follows:

\begin{equation}
 k \cdot \frac{P_{ETH}}{(1+\gamma)} P_{DAI}= - (m + b - \alpha P_{ETH}) P_{DAI} +  (b + c)
 \end{equation}

This yields the following equation for the price of DAI as a function of $\gamma$, $P_{ETH}$ and the belief parameter $b$:
\begin{equation}
    P_{DAI} = \frac{b+c}{b+m+P_{ETH}(\frac{k}{1+\gamma}-\alpha)}
\end{equation}

We can see that as $b \to \infty$, $P_{DAI} = 1$. Thus this model suggests that if buyers of DAI believe strongly that the price is close to one, they will shape the price accordingly. On the other hand, as $b\to 0$, $P_{DAI}=\frac{c}{m+P_{ETH}(\frac{k}{1+\gamma}-\alpha)}$, which may not converge at $1$.

We will see in the following section that empirically there is no evidence for a strong dependence of $P_{DAI}$ on $P_{ETH}$. In the model that we have presented above, this would happen if one of the following two conditions happens:
\begin{itemize}
    \item $k$ and $\alpha$ are both relatively small, which means that the dependence of DAI supply and demand on the price of ETH is weak to begin with.
    \item Or, $\frac{k}{(1+\gamma)} \approx \alpha$, which means that the two effects of relationship between supply and ETH price as well as the relationship between demand and ETH price more or less cancel each other out. 
\end{itemize}


\section{How collateral price impacts derivative price} \label{sec6}
\subsection{Empirical evidence}

\begin{figure}[h!]
    \centering
    \begin{minipage}{0.45\linewidth}
        \centering
        \includegraphics[trim={0 0 0 1cm}, clip, width=0.9\linewidth]{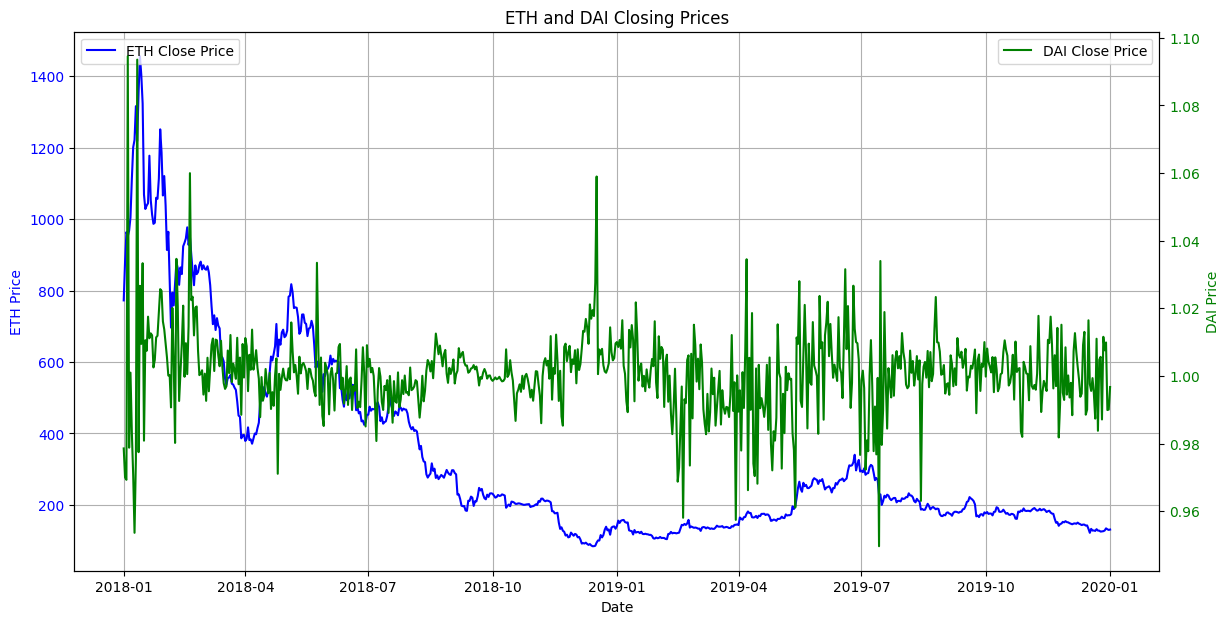}
        \caption{Real data of DAI price vs ETH price}
        \label{fig:label1}
    \end{minipage}\hfill 
    \begin{minipage}{0.45\linewidth}
        \centering
        \includegraphics[trim={0 0 0 1cm}, clip,width=0.9\linewidth]{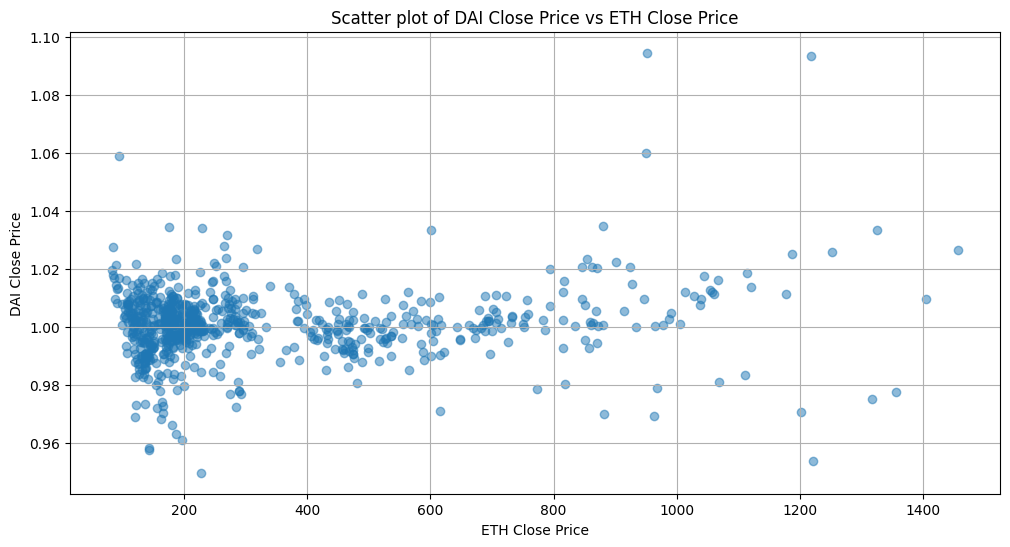}
        \caption{Scatter Plot of DAI price vs ETH price}
        \label{fig:label2}
    \end{minipage}
\end{figure}

Figure-\ref{fig:label1} demonstrates the closing prices of both ETH and DAI over the two-year period from 2018 to 2020. Ethereum (ETH) exhibited significant volatility, with its price ranging from approximately \$84 to \$1456. The average closing price for ETH during this period was about \$332.52. On the other hand, DAI demonstrated its characteristic stability. As a stablecoin, its price hovered around the \$1 peg, with minor fluctuations. The average closing price for DAI was approximately \$1.0007. To further understand the relationship between ETH and DAI, we calculated the correlation coefficient between their closing prices, which came out as 0.1336. The calculated correlation coefficient of 0.1336 suggests a weak positive linear relationship between the closing prices of ETH and DAI. The correlation coefficient being close to 0 is weak enough to draw any meaningful conclusions. 

\begin{table}[h]
\begin{center}
\begin{tabular}{|c|c|c|}
\hline
\textbf{Statistic} &  \textbf{ETH} & \textbf{DAI} \\ \hline
    mean &  332.521782 &   1.000718 \\ \hline
    std &  263.910774 &   0.012285 \\ \hline
    max & 1456.138411 &   1.094566 \\ \hline
    min &   84.248953 &   0.949661 \\ \hline
    25\% &  160.786182 &   0.995545 \\ \hline
    50\% &  211.537433 &   1.000806 \\ \hline
    75\% &  451.533303 &   1.006330 \\ \hline
\end{tabular}
\caption{Statistical values for DAI and ETH closing price}
\label{tab:eth_dai_stats}
\end{center}
\end{table}

Figure-\ref{fig:label2} illustrates the scatter plot of DAI vs ETH closing prices. We observe there's a high density of points around the DAI price of \$1, confirming its stable nature. This dense cluster is spread across various ETH prices, which is consistent with the observation that DAI maintains its peg to around \$1 regardless of ETH's price movements. The scatter plot reaffirms the distinct characteristics of the two assets. DAI, being a stablecoin, predominantly hovers around the \$1 mark, irrespective of the volatility and price changes exhibited by ETH. This scatter plot also aligns with the previously computed weak correlation coefficient, indicating that the movements in DAI prices have a weak linear relationship with ETH's price movements.

\subsection{Simulation results}

\begin{figure}[h!]
    \begin{minipage}{0.45\linewidth}
        \centering
        \includegraphics[trim={0 0 0 .8cm}, clip,width=0.9\linewidth]{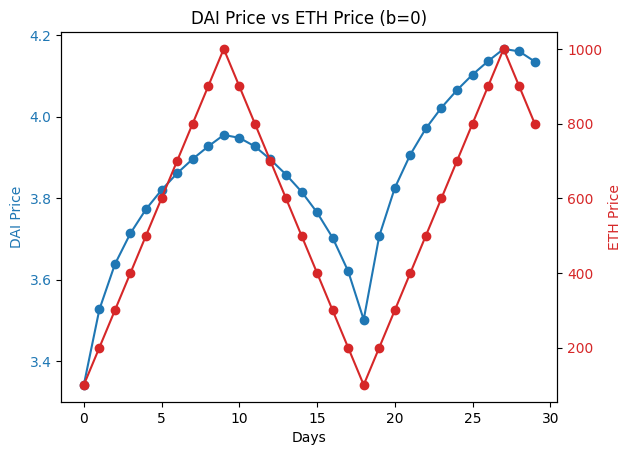}
        \caption{DAI price vs ETH price (b=0)}
        \label{fig:label5}
    \end{minipage}\hfill
    \begin{minipage}{0.45\linewidth}
        \centering
        \includegraphics[trim={0 0 0 .8cm}, clip,width=0.85\linewidth]{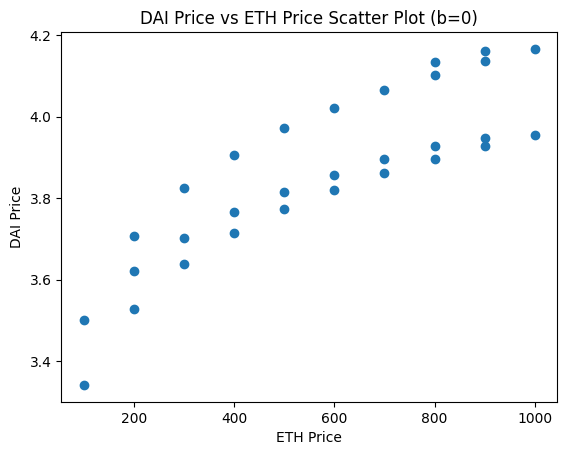}
        \caption{Scatterplot of DAI price vs ETH price (b=0)}
        \label{fig:label6}
    \end{minipage}
\end{figure}

\begin{figure}[h!]
    \begin{minipage}{0.45\linewidth}
        \centering
        \includegraphics[trim={0 0 0 .8cm}, clip, width=0.9\linewidth]{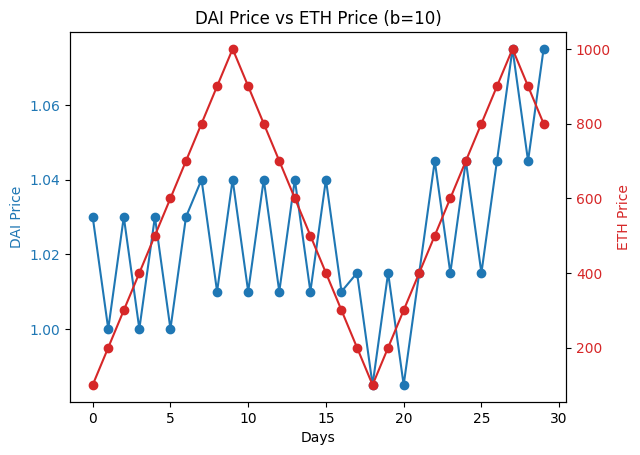}
        \caption{DAI price VS ETH price (b=10)}
        \label{fig:label3}
    \end{minipage}\hfill
    \begin{minipage}{0.45\linewidth}
        \centering
        \includegraphics[trim={0 0 0 .8cm}, clip, width=0.85\linewidth]{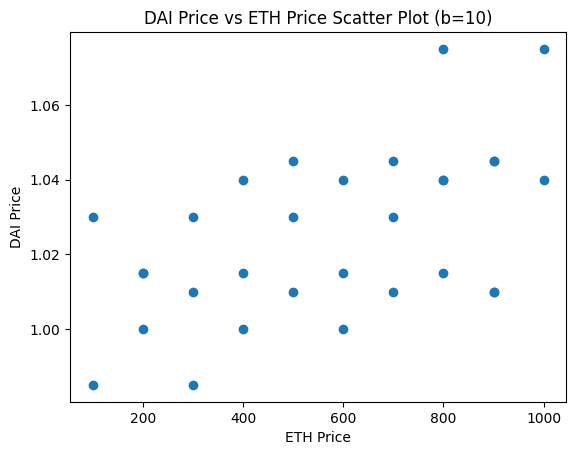}
        \caption{Scatterplot of DAI price vs ETH price (b=10)}
        \label{fig:label4}
    \end{minipage}
\end{figure}

Figures-\ref{fig:label5} \& \ref{fig:label3} demonstrates the predicted value of the price of DAI through simulation models. The DAISIM simulation model initially outputs DAI prices ranging from approximately \$3.34 to \$4.17. This shows that the predicted DAI price significantly deviates from its pegged value of \$1. This variation is attributed to the model's lack of consideration for market beliefs and narratives. However, with the integration of a belief parameter based on the hypothesis of DAI's peg to \$1, there is a substantial improvement in price stability. The modified model generates DAI prices that fluctuate closely around the \$1 mark with minimal deviations.

This improvement is also evidenced by examining the correlation coefficients between DAI and ETH prices in both models as depicted in Figures-\ref{fig:label6} \& \ref{fig:label4}. Initially, the correlation coefficient stands at 0.86, indicating a strong positive correlation. This suggests that DAI prices are significantly influenced by ETH's price movements, contrary to the observed behavior of DAI in the real world. However, after the introduction of the belief parameter, the correlation coefficient decreases to 0.54. The lower correlation more accurately mirrors the real-world expectation that DAI's price remains largely independent of ETH's market fluctuations. This observation reinforces the significance of integrating realistic market factors into financial simulation models.

\section{Risk Factors for Stable Derivatives} \label{sec7}
Stable derivatives represent stability against the volatility of cryptocurrencies in DeFi. Although these stable-coins are engineered to maintain peg with traditional currencies or commodities, they face various risk factors that can destabilize their value. 
\subsection{Underlying Blockchain}
The stability and operation of decentralized applications or protocols largely depend on the foundation provided by the underlying blockchain. If the blockchain experiences issues such as poor performance, delayed transactions, or security breaches, this impacts all applications built on it. A practical example is the Mirror Protocol's shutdown, which stemmed from Oracle issues on the Terra blockchain. Challenges faced by a blockchain can ripple through to DeFi protocols, diminishing user trust and reducing system stability.


\subsection{Oracle}
Oracles play a critical role in connecting blockchain systems with real-world data. They are responsible for delivering precise price data, ensuring proper collateralization within the system. However, they aren't exempt from risks. Events like pricing inaccuracies, unexpected system behaviors, or targeted attacks can result in oracles providing wrong data. This misinformation can set off unwarranted liquidations or other adverse system actions.

\subsection{Smart Contract Bugs or Hacks}
Smart contracts are core to DeFi protocol operations. Like other softwares, they are also prone to vulnerabilities. Threat actors continuously search for these vulnerabilities in the smart contract code to exploit. In the worst case, an attack on the infrastructure could result in a total loss of decentralized assets held as collateral.

\subsection{Failure of Centralized Infrastructure}
Despite DeFi's central tenet being decentralization, some aspects still rely on centralized systems, particularly in early stages. DeFi platforms sometimes lean on centralized systems or foundations for operations, management, and governance. Any malfunction or regulatory issues of these centralized parts can pose a considerable threat to the protocol. Issues could range from access barriers in specific countries, failure in front-end user interfaces, or legal complications for the overseeing foundation, all of which can affect the protocol's seamless operation and expansion.

\subsection{Debt Ceiling}
Considering single collateral DAI, the debt ceiling sets a maximum threshold for the amount of DAI that can be minted against a specified collateral, commonly ETH. Establishing this limit is a strategic management scheme to restrict the available supply of DAI in the DeFi market. However, the complex dynamics of the DeFi market can transform this into a potential risk factor for the stable derivative. When the amount of DAI generated against the single collateral reaches its debt ceiling, the system prohibits any further generation of DAI. As a result, users are compelled to acquire DAI exclusively from the open market. Such constraints can exert upward pressure on DAI's price, pushing it beyond its intended \$1 peg. This situation underscores the risk of a stable derivative's price surpassing its designated peg.

\section{Conclusions} \label{sec8}
We introduced a belief parameter in the modeling of the DAI stablecoin to effectively capture market sentiments regarding its valuation and stability. We constructed a mathematical that quantitatively depicts the stability mechanism of DAI in response to ETH price fluctuations. Additionally, we conducted an extensive risk analysis for stable derivatives, examining the impact of factors such as oracle reliability, debt ceiling, and smart contract vulnerabilities on their stability and operation. For future research, we plan to refine the model to reduce the correlation between DAI and ETH prices and expand the model to incorporate multi-collateral DAI in order to provide deeper insights into its stability and risk factors in a more diverse economic environment.

\label{Bibliography}

\bibliographystyle{IEEEtran}
\bibliography{references}

\begin{thebibliography}{10}
\providecommand{\url}[1]{#1}
\csname url@samestyle\endcsname
\providecommand{\newblock}{\relax}
\providecommand{\bibinfo}[2]{#2}
\providecommand{\BIBentrySTDinterwordspacing}{\spaceskip=0pt\relax}
\providecommand{\BIBentryALTinterwordstretchfactor}{4}
\providecommand{\BIBentryALTinterwordspacing}{\spaceskip=\fontdimen2\font plus
\BIBentryALTinterwordstretchfactor\fontdimen3\font minus \fontdimen4\font\relax}
\providecommand{\BIBforeignlanguage}[2]{{%
\expandafter\ifx\csname l@#1\endcsname\relax
\typeout{** WARNING: IEEEtran.bst: No hyphenation pattern has been}%
\typeout{** loaded for the language `#1'. Using the pattern for}%
\typeout{** the default language instead.}%
\else
\language=\csname l@#1\endcsname
\fi
#2}}
\providecommand{\BIBdecl}{\relax}
\BIBdecl

\bibitem{kolodziejczyk2020stablecoin}
H.~Ko{\l}odziejczyk and K.~Jarno, ``Stablecoin--the stable cryptocurrency,'' 2020.

\bibitem{ante2021influence}
L.~Ante, I.~Fiedler, and E.~Strehle, ``The influence of stablecoin issuances on cryptocurrency markets,'' \emph{Finance Research Letters}, vol.~41, p. 101867, 2021.

\bibitem{team2020maker}
M.~Team, ``The maker protocol: Makerdao’s multi-collateral dai (mcd) system,'' \emph{White paper}, 2020.

\bibitem{lyons2023keeps}
R.~K. Lyons and G.~Viswanath-Natraj, ``What keeps stablecoins stable?'' \emph{Journal of International Money and Finance}, vol. 131, p. 102777, 2023.

\bibitem{mita2019stablecoin}
M.~Mita, K.~Ito, S.~Ohsawa, and H.~Tanaka, ``What is stablecoin?: A survey on price stabilization mechanisms for decentralized payment systems,'' in \emph{2019 8th International Congress on Advanced Applied Informatics (IIAI-AAI)}.\hskip 1em plus 0.5em minus 0.4em\relax IEEE, 2019, pp. 60--66.

\bibitem{klages2020stablecoins}
A.~Klages-Mundt, D.~Harz, L.~Gudgeon, J.-Y. Liu, and A.~Minca, ``Stablecoins 2.0: Economic foundations and risk-based models,'' in \emph{Proceedings of the 2nd ACM Conference on Advances in Financial Technologies}, 2020, pp. 59--79.

\bibitem{gudgeon2020decentralized}
L.~Gudgeon, D.~Perez, D.~Harz, B.~Livshits, and A.~Gervais, ``The decentralized financial crisis,'' in \emph{2020 crypto valley conference on blockchain technology (CVCBT)}.\hskip 1em plus 0.5em minus 0.4em\relax IEEE, 2020, pp. 1--15.

\bibitem{kothari2018simulating}
T.~Kothari and W.~C. Gu, ``Simulating stablecoin systems with latent market confidence index,'' \emph{Available at SSRN 3508036}, 2018.

\bibitem{clark2020demystifying}
J.~Clark, D.~Demirag, and S.~Moosavi, ``Demystifying stablecoins: Cryptography meets monetary policy,'' \emph{Queue}, vol.~18, no.~1, pp. 39--60, 2020.

\bibitem{synthetix}
\BIBentryALTinterwordspacing
Synthetix, ``Synthetix: Decentralized synthetic assets,'' 2023. [Online]. Available: \url{https://www.synthetix.io}
\BIBentrySTDinterwordspacing

\bibitem{mirror}
\BIBentryALTinterwordspacing
M.~Protocol, ``Mirror protocol: Synthetic assets on the blockchain,'' 2023. [Online]. Available: \url{https://www.mirror.finance}
\BIBentrySTDinterwordspacing

\bibitem{bhat2021daisim}
S.~Bhat, A.~B. Kahya, B.~Krishnamachari, and R.~Kumar, ``Daisim: A computational simulator for the makerdao stablecoin,'' in \emph{4th International Symposium on Foundations and Applications of Blockchain 2021 (FAB 2021)}.\hskip 1em plus 0.5em minus 0.4em\relax Schloss Dagstuhl-Leibniz-Zentrum f{\"u}r Informatik, 2021.

\bibitem{DAISIM2}
\BIBentryALTinterwordspacing
ANRGUSC, ``Daisim 2.0,'' 2024. [Online]. Available: \url{https://github.com/ANRGUSC/DAISIM}
\BIBentrySTDinterwordspacing

\end{thebibliography}

\end{document}